# Combined funnel, concentrator, and particle valve functional element for magnetophoretic bead transport based on engineered magnetic domain patterns

*Rico Huhnstock*[1,2], *Lukas Paetzold*[1,2], *Maximilian Merkel*[1,2], *Piotr Kuświk*[3] *and Arno Ehresmann*[1,2]

1) Institute of Physics and Center for Interdisciplinary Nanostructure Science and Technology (CINSaT), University of Kassel, Heinrich-Plett-Str. 40, D-34132 Kassel, Germany

2) Artificial Intelligence Methods for Experiment Design (AIM-ED), Joint Lab of Helmholtzzentrum für Materialien und Energie, Berlin (HZB) and University of Kassel, Hahn-Meitner-Platz 1, D-14109 Berlin, Germany

3) Institute of Molecular Physics, Polish Academy of Sciences, M. Smoluchowskiego 17, Poznań, 60-179, Poland

*Corresponding author: rico.huhnstock@physik.uni-kassel.de

ABSTRACT

Controlled actuation of superparamagnetic beads (SPBs) within a microfluidic environment using tailored dynamic magnetic field landscapes (MFLs) is a potent approach for the realization of point-of-care diagnostics within Lab-on-a-chip (LOC) systems. Making use of an engineered magnetic domain pattern as the MFL source, a functional LOC-element with combined magnetophoretic "funnel", concentrator, and "valve" functions for micron-sized SPBs is presented. A parallel-stripe domain pattern design with periodically increasing/decreasing stripe lengths has been fabricated in a topographically flat continuous exchange biased (EB) thin film system by ion bombardment induced magnetic patterning (IBMP). It is demonstrated that, upon application of external magnetic field pulses, a fully reversible concentration of SPBs at the domain pattern's focal point occurs. In addition, it is shown that this functionality may be used as an SPB "funnel", allowing only a maximum number of particles to pass through the focal point. Adjusting the pulse time length, the focal point can be clogged up for incoming SPBs, resembling an on/off switchable particle "valve". The observations are supported by quantitative theoretical force considerations.

*Introduction*

Superparamagnetic beads (SPBs) are discussed as central components of future point-of-care diagnostic and analytic devices [1, 2, 3], realized in micro-total-analysis-systems (µTAS) or Lab-on-a-chip (LOC) devices. [4, 5, 6] They are available in different sizes, compositions, and chemical surface functionalizations, enabling specific analyte binding and isolation from a screened fluid. [7, 8] SPB-based biodetection schemes discussed in literature rely on the particles' remotely-controllable actuation. A major strategy for actuation is the use of local field gradients within tailored magnetic stray field landscapes (MFLs) emerging from a chip substrate superposed by a dynamically varying external magnetic field. [9, 10, 11, 12, 13] These MFLs have been created by periodic arrays of micro-structured soft-/hardmagnetic elements [14, 15, 16, 9, 17] or by magnetic domains in full magnetic thin film systems [18, 19, 20, 10, 21], the latter either occurring naturally in ferromagnetic garnet films [18] or artificially tailored by magnetic patterning of continuous exchange bias (EB) [20, 10, 22, 23] or multilayer [24] thin film systems. Prominent examples of magnetic patterning techniques for EB thin film systems are ion bombardment induced magnetic patterning (IBMP) [23], thermally assisted scanning probe lithography [25] and laser-based direct-writing [26]. The MFL's steep magnetic field gradients between adjacent field minima and maxima (typically separated by a few μm) yield comparably high SPB steady-state transport velocities with more than 100 μm/s. [18, 12, 27]

An important functionality of dynamic MFLs suitable for LOC devices is their ability to guide SPBs towards a sensing position in order to increase analyte detection sensitivity [28], thus, emphasizing the need for a matching micromagnetic pattern within the underlying substrate. For instance, conducting micro-loops [29, 30], spiderweb-like [28], concentric cylinder [31] as well as periodic circular micromagnetic structures [32] have been utilized to controllably focus SPBs towards a designated on-chip area, either with the goal of detecting the particles [28, 29, 32] or reducing the interparticle distance for a potential analyte-induced particle aggregation [31]. However, lacking control over the number of SPBs arriving at the focus position is oftentimes a drawback: Typically, all particles subjected to the MFL above the micromagnetic pattern will be focused, resulting in fluctuating amounts. A defined number of particles would be beneficial when combining the focusing step with magnetoresistive particle detection [33, 29, 30] since the measured signal could be directly used for sensor calibration. In case of analyte-induced SPB aggregation, no regulation of the number of incoming particles would accordingly lead to no control over the aggregate sizes, which especially hinders the quantification of analyte concentrations in such detection assays. Additionally, analyte detection based on particle aggregation typically combines two steps: First, bead aggregation has been induced by a permanent magnet prior to on-chip handling, accumulating beads close to the magnet and, thus, enhancing the probability of binding events [32, 34, 35]. Then, in a second step, analyte-bridged SPBs have been separated from single ones by non-linear magnetophoresis in dynamically varying MFLs [32, 34].

Overcoming these downsides, we have designed in this work a functional element with tailored MFL, which, in combination with a superposed dynamically varying external field fulfills two necessities: 1.) Limitation of the amount of transported SPBs towards a focusing region and 2.) combination of the two steps of (I) bead approach for analyte-induced binding and (II) separation of beads after their approach in a continuous motion cycle. Starting from a prototypical IBMP-engineered parallel-stripe domain pattern with periodically alternating head-to-head (hh)/ tail-to-tail (tt) magnetization configurations [20, 10], the stripe lengths have been gradually varied for the presented study, creating a "magnetophoretic funnel" for laterally transported SPBs. Decreasing the stripe domain length, while keeping the width constant, shortens the trapping sites for SPBs in one dimension, consequently bringing the particles closer to each other (Figure 1a). Lowered interparticle distances are expected in this case because a reduced stray field strength for shorter domain walls (DWs) weakens the dipolar repulsion between single SPBs. Additionally, fringe fields at the upper/lower boundaries of magnetic stripe domains in *y*-direction should further stabilize SPB formations with reduced interparticle distances. As visualized by Figure 1a, this has two effects: On the one hand, the relation of SPB size and DW length allows only for a limited number of SPBs passing through during lateral particle transport. On the other hand, the accompanied proximity of SPBs to each other increases the probability of analyte bridge formation between functionalized SPBs when analytes (represented by yellow squares in Figure 1a) are present. Demonstrating our capability to reversibly increase the interparticle distance and therefore dissolve SPB aggregates with no analyte bridges, the DW length is gradually increased beyond the focal point of the pattern. It is expected that SPBs will be well separated again in this region since dipolar repulsion between SPBs is becoming more pronounced due to parallelly aligned magnetic moments [10] and increasing stray field strengths. As a proof-of-principle of the discussed advantages, we experimentally test the lateral transport of SPBs without surface functionalization, correlating the SPB motion dynamics observed by an optical bright-field microscope with the tailored magnetic domain pattern imaged via magnetic force microscopy (MFM) as well as with numerical simulations for the resulting MFL and the acting forces on the particles. By varying the time length of externally applied magnetic field pulses, we will additionally demonstrate that beyond a critical threshold the "magnetophoretic funnel" can be jammed for incoming SPBs, adding a particle "valve" functionality to the utilized domain pattern.

*Results*

*Magnetic domain pattern and magnetic stray field landscape*

Magnetic parallel-stripe domain patterns with periodically increasing and decreasing stripe lengths were fabricated within EB thin film systems via IBMP (see Figure S1 in the Supporting Information for an image of the utilized resist structure). The stripes of 5 µm width and remanent in-plane magnetization along their short axes are hereby embedded in a monodomain phase of opposite magnetization direction in remanence. The length of the stripes was varied between 5 µm and 50 µm with an increment of 5 µm, i.e., the vertical distance between upper/lower edges of consecutive stripes amounts to 2.5 µm. To confirm the successful fabrication of the domain pattern and to investigate occurring DW types, MFM measurements were performed for an exemplary sample area of 80 µm × 80 µm with a tip elevation of 200 nm above the substrate. The result is shown in Figure 1b, with the MFM phase signal represented by a pseudocolor. Here, DWs between stripe domains and the surrounding monodomain environment are clearly visible as regions with the darkest (right boundary of a stripe) and lightest (left boundary of a stripe) phase signal contrast. As the stripe domains are antiparallelly magnetized with respect to their environment (indicated by blue arrows) three different DW configurations can be identified: hh, tt, and side-by-side (ss). Each of the corresponding DWs carries a different magnetic charge distribution profile, resulting in different strengths for the emerging magnetic stray fields [36].

To study the theoretical MFL emerging from this domain pattern, micromagnetic simulations, using the MuMax3 software package [37], were performed for the sample area marked by the black rectangle in Figure 1b. The used simulation parameters and dimensions can be found in the Methods section. From the obtained distribution of magnetic moments $\vec{m}$, magnetic stray fields at position $\vec{r}(x, y, z)$ were calculated according to the dipole approximation [38]

$$\vec{H}(\vec{r}) = \frac{1}{4\pi} \cdot \sum_i \frac{3 \cdot (\vec{R} \cdot \vec{m}_i) \cdot \vec{R}}{|\vec{R}|^5} - \frac{\vec{m}_i}{|\vec{R}|^3}. \tag{1}$$

Within this formula, $\vec{R} = \vec{r} - \vec{r}_i$ denotes the distance vector between spatial position $\vec{r}$ and dipole position $\vec{r}_i$. The different MFL components $H_z$, $H_x$, and $H_y$ in z-, x-, and y-direction were computed at a distance of 1600 nm above the substrate surface and are depicted as pseudocolor plots in Figure 1c.

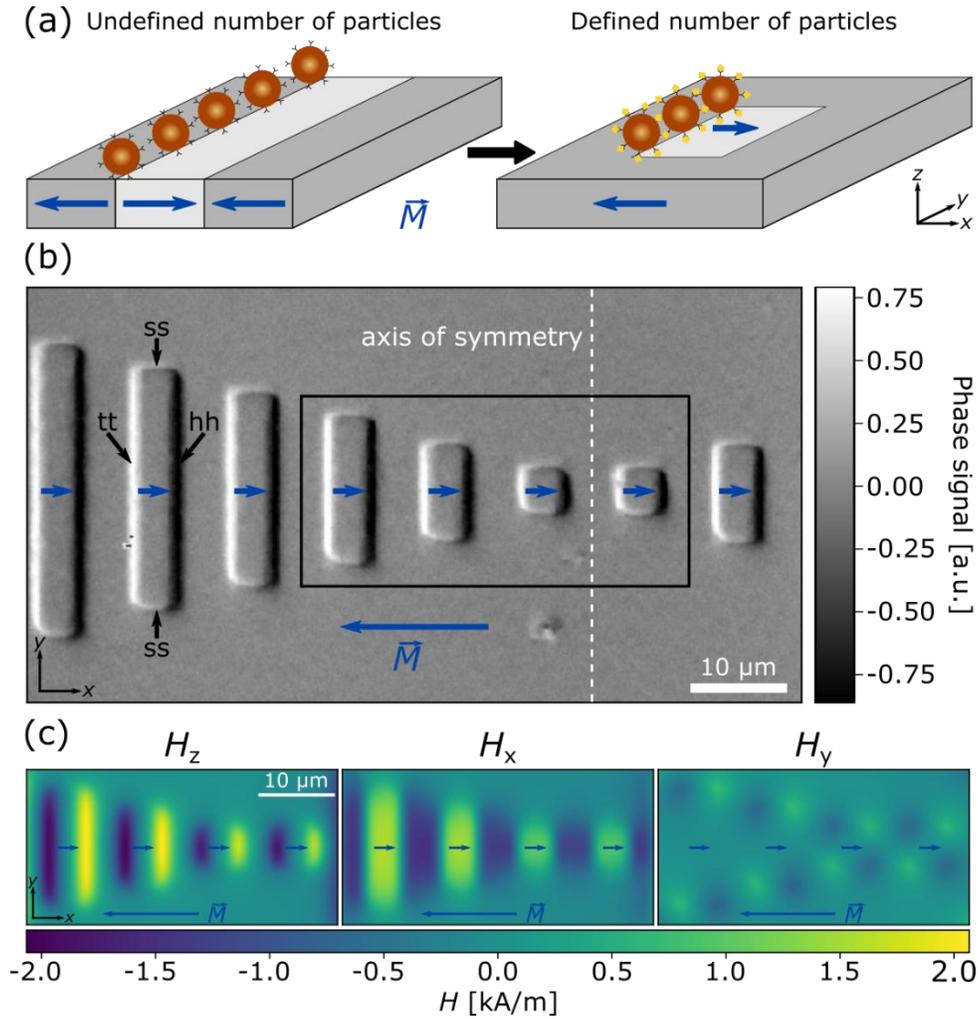

Figure 1: Designed magnetic parallel-stripe domain pattern with gradually increasing/decreasing length for controlled funneling and accumulation of SPBs. (a) Schematic concept for limiting the number of accumulated SPBs and subsequent analyte detection facilitated by the designed domain pattern. The probability of analyte bridge formation between SPBs and thereby particle aggregation is expected to be increased when stripe domain lengths are decreased. (b) Phase contrast MFM image at an elevation of $z = 200$ nm above the substrate. Light and dark grey levels mark positions of DWs. Grey level changes across DWs indicate head-to-head (hh), tail-to-tail (tt), or side-by-side (ss) DWs [23, 36]. The axis of symmetry is indicated at which the domain pattern is mirrored. (c) Simulated MFL at an elevation $z = 1600$ nm above the magnetic substrate for the region of interest signified by the black frame in (b). Shown are the magnetic field components $H_z$, $H_x$, and $H_y$ as a function of position, respectively.

The elevation of 1600 nm was chosen due to the radius of investigated SPBs being 1400 nm and the thickness of a Poly(methyl methacrylate) (PMMA) spacing layer on top of the magnetic thin film system being 200 nm, thus, the MFL was simulated at the approximated position of physical particle centers above the substrate, which is expected to coincide also with the magnetic center of the SPBs. Computed stray fields can, therefore, be used to estimate the magnitude of magnetic forces acting on the studied SPBs. As it can be seen from Figure 1c, maximum magnitudes for $|H_z|$ are found directly at the positions of DWs with hh and tt magnetization configuration of adjacent domains. On the contrary, maximum $|H_x|$ occurs between hh and tt DWs, i.e., either above a stripe domain or above the gap between two stripe domains. As a common feature for $H_z$ and $H_x$ distributions, the magnetic field magnitude is predicted to be lowered above the smallest stripe domains of 5 µm length compared to stripe domains with larger lengths. For instance, maximum $|H_z|$ is 2.05 kA/m for DWs of 10 µm length and 1.74 kA/m for 5 µm DWs according to the simulation results. This trend proves to be vital for finding a physical explanation for the herein-discussed SPB concentrator and "valve" functionalities of the magnetic domain pattern (see Discussion section for a more detailed theoretical analysis). Examining the computed distribution of $H_y$, alternating maxima/minima can be identified at the corners of each stripe domain. These "fringe fields" are expected to play a crucial role in the funneling and concentration of SPBs described below. Compared to $H_z$ however, the maximum magnitude $|H_y|$ is approximately only 30 % of maximum $|H_z|$. Judging solely from these magnetic stray field strengths, SPBs situated above the substrate should be primarily attracted towards the positions of DWs with hh and tt magnetization configurations.

*SPB motion dynamics*

An aqueous dispersion of SPBs (Dynabeads M-270) with a diameter of 2.8 µm was placed on top of the magnetically patterned substrate. Similar to previous works [10, 12, 13, 27], a one-directional, stepwise transport of the particles was initialized by applying a sequence of external trapezoidal magnetic field pulses in *z*- and *x*-direction with magnitudes of $\mu_0 \cdot H_{z,max} = \mu_0 \cdot H_{x,max} = 1$ mT ($\mu_0$ being the vacuum permeability). This strength of the external field pulses was chosen in order to be in the range of maximum MFL magnitudes and to simultaneously avoid remagnetization of the magnetic substrate. The field direction is hereby alternating periodically and both sequences for *z*- and *x*-direction are separated by a phase shift of $\pi/2$. Recording the ensuing SPB motion with a high speed camera (1000 frames per second) attached to an optical bright-field microscope, exemplary snapshots are shown in Figure 2 for experimental times of *t* = 0 s (a), *t* = 1.0 s (b), and *t* = 2.5 s (c).

As indicated by the red arrows, the particles are moving from the left to the right of the chosen field of view and their positions closely resemble the decreasing and subsequently increasing length of magnetic stripe domains in the underlying substrate. It is observed that formations of SPBs are compacted with respect to the *y*-dimension upon moving to the sample position of smallest stripe domain length (here denoted as focal point), with a spreading of the particles occurring once they moved through the focal point and reached stripe domains of increasing length. At the beginning of the experiment, SPBs are situated within vertical rows on the left side from the focal point (Figure 2a). This is comparable to results for magnetic parallel-stripe domain patterns where the stripe length is equal to the substrate size [10]. When interacting with the MFL that is superposed with the external field, minima for the SPBs' potential energy are present above the DWs with either hh or tt magnetization configuration [10], leading to capture of the particles at these positions.

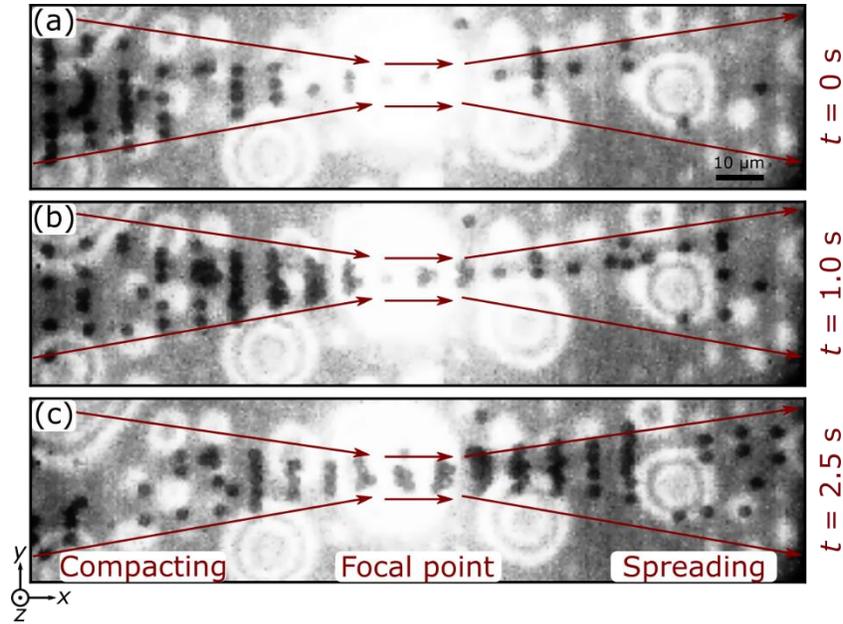

Figure 2: Directed transport of SPBs on top of a magnetic parallel-stripe domain pattern with gradually decreasing/increasing stripe lengths. The SPB motion was initiated by transforming the emerging MFL with external magnetic field pulses in $z$- and $x$-direction. Spatial positions and formations of the particles (black spots) were captured via optical bright-field microscopy and displayed exemplarily for experimental times of $t = 0$ s (a), $t = 1.0$ s (b), and $t = 2.5$ s (c).

Shifting these potential energy minima via the periodic external magnetic field pulses results ultimately in the movement of the SPBs along the $x$-direction. As the magnetic moments of the SPBs are aligned along the effective magnetic field and therefore in a parallel configuration [10], particles are repelling each other when being situated within the observed row formation. This situation changes when the SPBs move towards the focal point: If more than two particles arrive simultaneously at this position, a transition to a clustered formation is observed where particles are arranged in a hexagonal-like lattice. The transformation is already initiated before the focal point (see Figure 2b) since single particles are ejected to the sides of a row formation when the physical space above a DW is not sufficient for all SPBs present. Interestingly, the arrangement of SPBs is reversibly transformed back to the row formation upon reaching the sample area with increasing stripe domain lengths. As a first implication, the observed reversible and stepwise switching between clustered and row formation of SPBs means that the interparticle distance is tunable, demonstrating the suitability of the presented approach for controlled establishment of molecular bridges between SPBs for analyte detection.

Upon studying the influence of the external magnetic field pulse duration on the SPBs' transport behavior, another striking feature of the employed magnetic domain pattern was discovered. As known from literature for the utilized magnetic particle transport approach of a dynamically transformed potential energy landscape, SPBs are transiting into a non-linear motion regime ("phase-slipping") if the frequency of an external rotational magnetic field is exceeding a certain critical frequency [9]. Similarly, SPBs moving above a parallel-stripe domain pattern lose their transportability if the duration of the external magnetic field pulses is too short [10]. For the here investigated gradually modified stripe domain pattern, a combination of both linear and non-linear particle transport was observed depending on the sample position: At a critical external pulse duration, SPBs present within the focal point of the pattern performed an oscillating motion, "clogging up" the transport track for all approaching particles that are still moving directionally. This observation is visualized in Figure 3 by showing four microscope image snapshots (a-d) of an experiment, where the external magnetic field pulse duration was chosen to be $T = 40$ ms. As for the previously discussed experiment, the magnitudes of the field pulses stayed at $\mu_0 \cdot H_{z,max} = \mu_0 \cdot H_{x,max} = 1$ mT.

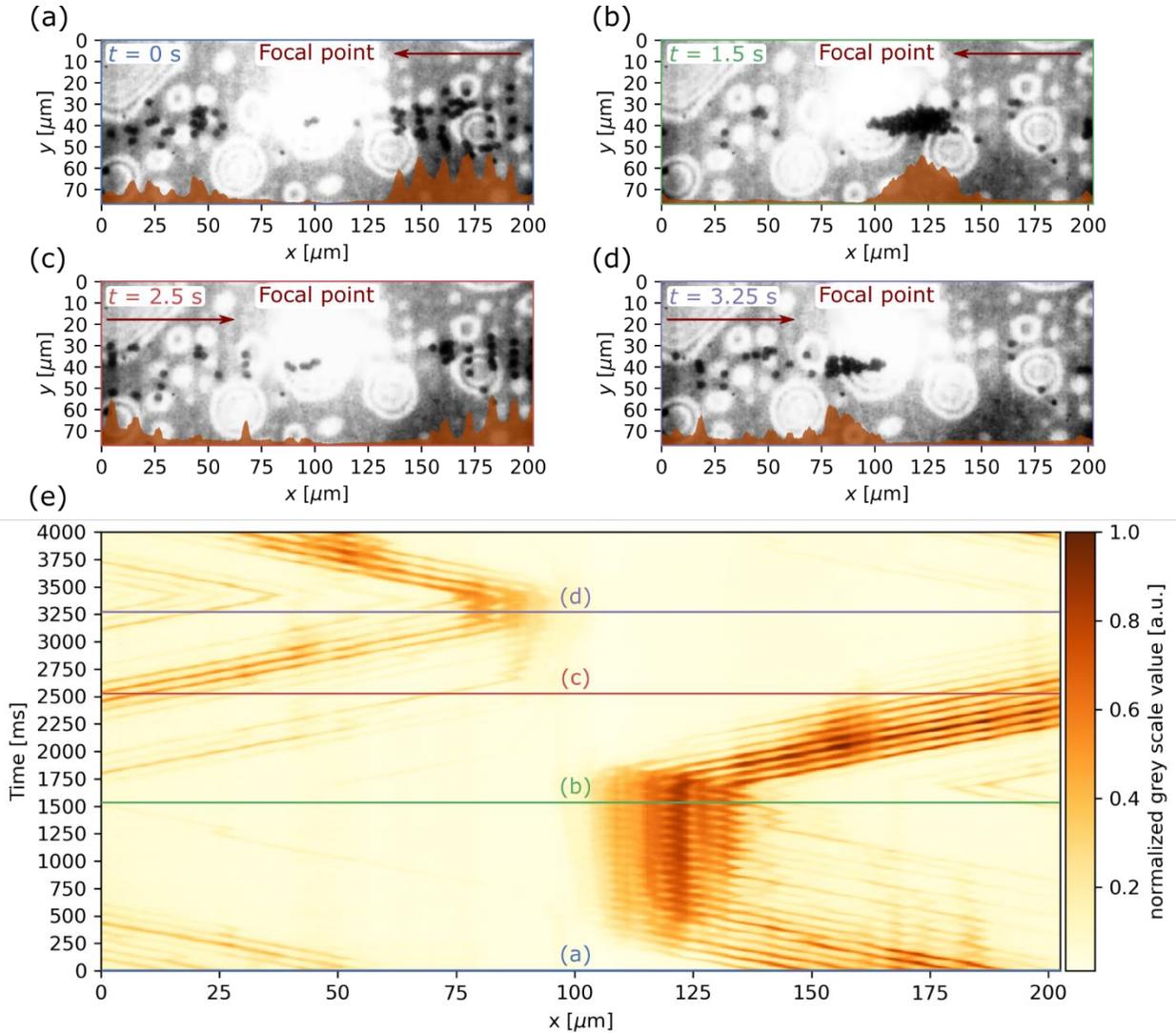

Figure 3: Motion behavior of SPBs above magnetic parallel-stripe domains with gradually increasing/decreasing stripe length when applying comparably short magnetic field pulses of $T = 40$ ms. Shown are four exemplary snapshots (a-d) from a transport video recording at different experimental times $t$ together with lateral intensity profiles for each frame (brown). Within these profiles grey scale values of the image were averaged along the $y$-axis and displayed as a function of $x$-position. SPBs are approaching the focal point of the domain pattern (located in the images at 100 μm) from the right side (a) and are entirely blocked at this position (b). Analogous observations were made when SPBs are approaching the focal point from the opposite side (c,d). Motion dynamics of the particles visualized by plotting the frame-by-frame lateral intensity profiles of the video recording as a function of time (e). Colored horizontal lines indicate the times at which frames (a-d) were taken.

Figure 3(a) shows the start of an experiment, where at $t = 0$ s the SPBs are moving from the right side towards the region of smallest stripe domain length (focal point, located at $x = 100$ μm). For quantification of average SPB positions, grey scale values of the image were averaged along the $y$-axis and are shown as a function of $x$-position (brown-filled curve) within the chosen video frame. Peaks in these lateral intensity profiles represent particle rows forming along the $y$-axis. When reaching the focal point, the SPBs do not move further but perform oscillating movements around a fixed position. Thus, all other SPBs approaching the focal point were not able to physically pass the stuck particles, leading to highly spatially concentrated aggregates. In the experiment displayed in Figure 3, this has been observed for $t = 1.5$ s (Figure 3b). This behavior is expected to be closely related to a decrease in MFL strength for the smallest stripe domain length (further elaboration on this mechanism in the Discussion section). Upon inverting the phase relation between the magnetic field pulses in $z$- and $x$-direction, the SPB transport direction was changed. Now all particles that were previously blocked have been transported away from the focal point. At the same time, new SPBs are moving toward this position from the left side (Figure 3c). At $t = 3.25$ s, also those particles are clogging at the focal point (Figure 3d). In Figure 3e the lateral intensity profiles are plotted as functions of time for the conducted experiment. The progression of each brown line indicates the motion of a single SPB row. Starting from $t = 0$ ms towards $t \approx 500$ ms, diagonal lines are indicating the directed transport of SPBs towards the focal point of the underlying domain pattern. With increasing time, SPBs start to oscillate around a constant $x$-position with a more blurred distribution of intensity indicating the breakup of distinct particle row formations. At $t \approx 1750$ ms, the phase relation for the external field sequence has been inverted, leading again to diagonal intensity lines that represent the SPB motion from the left to the right. Again, particles perform oscillating movements upon being in proximity to the focal point, most prominently visible between $t \approx 3000$ ms and $t \approx 3500$ ms. Another phase inversion for the external field sequence results once more in a directed transport of SPBs away respectively towards the focal point.

*Discussion*

The functional element for magnetophoretic bead transport introduced here is based on a clever design of magnetic domains and combines three functionalities: 1.) A particle "valve", 2.) a particle concentrator, and 3.) a particle "funnel" for transferring a restricted number of particles per transport step. The two latter are connected to an observed transformation of SPB rows into SPB clusters upon lateral transport toward the domain pattern's focal point. Inferring from the recorded microscope images, the distance between single SPBs is largely reduced for the clustered state so that the particles seem to physically touch each other. Compared to previous results for SPBs assembled above DWs of similar magnetic stripe patterns, this behavior is not intuitively understandable since magnetic moments of the particles should be aligned parallel to each other by the effective magnetic field, thereby leading to repulsive dipolar forces between particles [10]. When analyzing the row-to-cluster transformation, the interparticle distance within an SPB row is decreased with smaller stripe length until particles are ejected to the side bit by bit. It was hereby found that only cluster formations with a number of SPBs $n \leq 6$ were transported through the focal point. For a higher *n*, the surplus particles were either discarded to an SPB formation following behind or sometimes even completely dislodged from the MFL. This observation highlights the capability of the studied system to be used as a "magnetophoretic funnel": The amount of SPBs transported through the focal point of the pattern is limited by the relation of their sizes to the minimal DW length. For the conducted proof-of-principle experiment with a particle diameter of 2.8 μm and a minimal DW length of 5 μm, the maximum number of SPBs within a single cluster formation was found to be 6. For a similar investigation with the same SPBs being transported above a slightly different domain pattern with a minimal DW length of 2 μm (detailed results are not shown here as they would be beyond the scope of this work), the maximum number changed to 3. As our experiments show, the amount of SPBs present within the focal region is additionally influenced by the overall particle concentration. Depending on the number of particles within an approaching row formation, different cluster formations can be identified within the focal point of the domain pattern, which is exemplarily depicted for 2, 3, 4, 5, and 6 SPBs in Figure 4a.

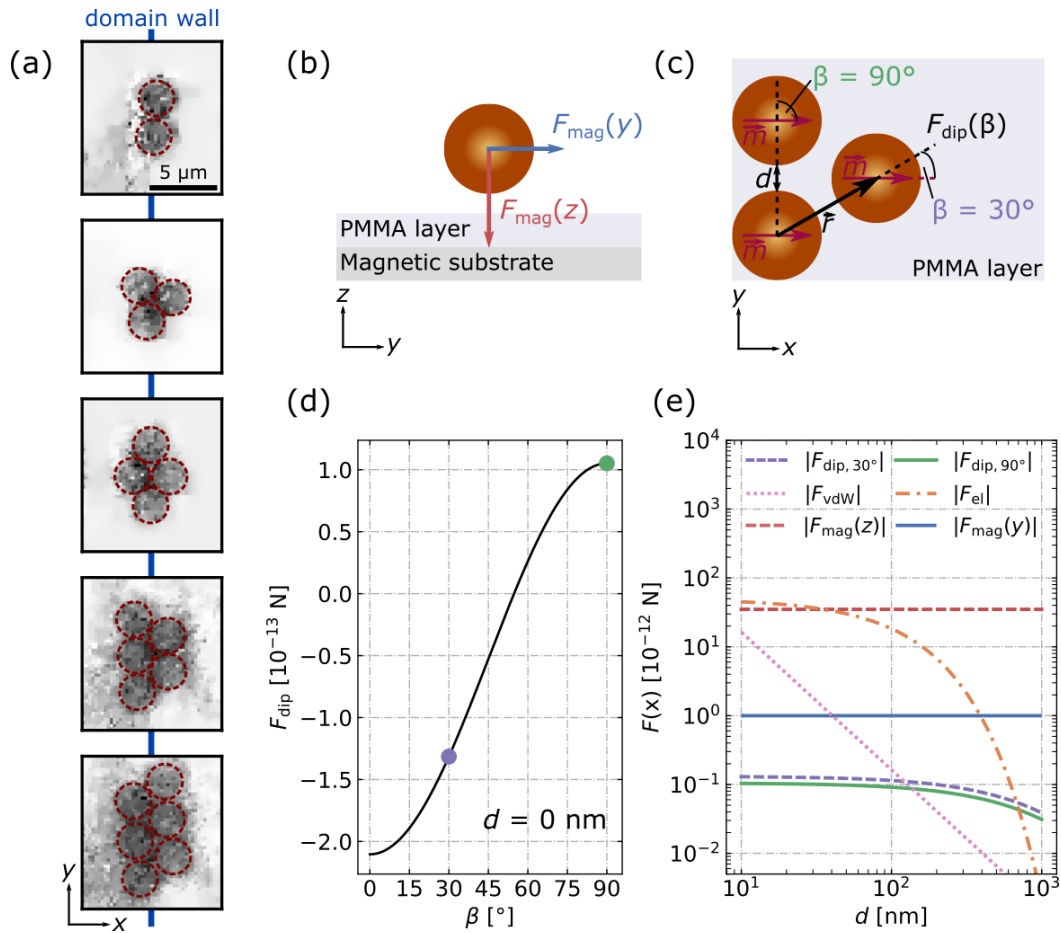

Figure 4: Qualitative characterization of SPB cluster formations observed during the conducted transport experiments and quantitative theoretical study of the most prominent forces acting on the particles. (a) Exemplary microscope images of SPB clusters forming over the focal point of the employed magnetic stripe domain pattern. The blue line indicates the approximated position of a DW in the underlying substrate. (b,c) Sketch of the involved forces that majorly guide the transport and clustering behavior of SPBs. Shown are the magnetic forces $F_{mag}(z)$ and $F_{mag}(y)$ exerted onto a single SPB by the MFL stemming from the substrate in $z$- and $y$-direction, respectively, (b) together with the dipolar force $F_{dip}(\beta)$ acting between two SPBs with magnetic moment $\vec{m}$ and separation distance $d$ (c). Here, the experimentally observed situations for the angles $\beta = 90°$ and $\beta = 30°$ are depicted. (d) Plot of calculated $F_{dip}$ as a function of $\beta$ for an interparticle distance of $d = 0$ nm. Highlighted are the angles 30° (purple circle) and 90° (green circle). (e) Plot of calculated forces acting on a single SPB situated above a DW as a function of interparticle distance $d$. Shown are the dipolar forces $F_{dip,30°}$ (purple dashed line) and $F_{dip,90°}$ (green solid line) for angles $\beta = 30°$ and $\beta = 90°$, respectively, the magnetic forces $F_{mag}(z)$ (red dashed line) and $F_{mag}(y)$ (blue solid line), the van der Waals force $F_{vdW}$ (pink dotted line), and the electrostatic force $F_{el}$ (orange dash dotted line).

Microscope images in Figure 4a show that SPB clusters align vertically along a DW in the underlying substrate, i.e., along the line of highest magnetic stray field strength. To further understand the physical mechanisms behind the observed cluster formation, acting forces that are expected to be of major influence were theoretically estimated. First, the magnetic force $\vec{F}_{\text{mag}}$ that is exerted onto a single SPB within the MFL is considered. It can be expressed by the gradient of the SPB's potential energy $U$ [10]:

$$\vec{F}_{\text{mag}}(x,y,z) = -\vec{\nabla} U(x,y,z) = -\mu_0 \cdot \left(\vec{m}_{\text{b}}(x,y,z) \cdot \vec{\nabla}\right) \cdot \vec{H}_{\text{eff}}(x,y,z), \qquad (2)$$

with $\vec{m}_{\text{b}}$ being the SPB's magnetic moment and $\vec{H}_{\text{eff}}$ being the effective magnetic field. The latter is a superposition of the locally present stray field $\vec{H}_{\text{MFL}}$ and the applied external magnetic field $\vec{H}_{\text{ext}}$. For simplification, we consider for this theoretical discussion a SPB that is located directly above a magnetic DW. As sketched in Figure 4b, the magnetic forces in z- and y-direction are of interest for the following reasons: 1.) the magnetic force $F_{\text{mag}}(z)$ is pulling SPBs towards a DW and 2.) the magnetic force $F_{\text{mag}}(y)$ exerted by the fringe fields sitting at the corners of a stripe domain (see Figure 1c) is additionally attracting SPBs towards these corners. The latter effect is expected to have a significant impact on the observed funneling of SPBs during lateral transport. For the interaction of SPBs with each other, we take the following forces into account: The electrostatic force $\vec{F}_{\text{el}}$, the van der Waals force $\vec{F}_{\text{vdW}}$, and the magnetostatic dipolar force $\vec{F}_{\text{dip}}$. For computing $F_{\text{el}}$ and $F_{\text{vdW}}$, the interaction of two spheres with equal radii $r = 1.4$ µm being surrounded by an aqueous medium is considered. According to the DLVO theory, $F_{\text{el}}$ and $F_{\text{vdW}}$ can be expressed as [39]:

$$F_{\text{el}} = 2\pi r \varepsilon_0 \varepsilon \kappa \psi_0^2 e^{-\kappa d}, \qquad (3)$$

$$F_{\text{vdW}} = -\frac{A \cdot r}{12 \cdot d^2}. \tag{4}$$

Here, $\varepsilon_0$ describes the electric field constant, $\varepsilon$ the relative permittivity, $\kappa$ the inverse Debye-Hückel double layer thickness, $\psi_0$ the SPB's surface potential, $A$ the Hamaker constant for the considered system, and $d$ the SPBs' surface-to-surface distance. For estimating $\vec{F}_{\text{dip}}$, the following assumptions were made: 1.) The effective magnetic field induces a fixed magnetic moment within the SPB that can be approximated as a point dipole located at the particle's center. 2.) SPBs are assumed to be positioned centrally above a DW with respect to the *x*- and *y*-direction, i.e., the influence of $H_{\text{MFL,x}}$ and $H_{\text{MFL,y}}$ is considered to be negligible. 3.) Only the projection of the SPBs' magnetic moments $\vec{m}$ onto the *x*-direction is taken into account, with all magnetic moments being aligned parallel to each other and being separated by the SPB-to-SPB-center distance $\vec{r}$ (see Figure 4c). Stemming from the magnetic dipole-dipole interaction energy, the dipolar force acting between two SPBs with an angle β between $\vec{m}$ and $\vec{r}$ is given by [40]

$$\vec{F}_{\text{dip}}(\beta) = \frac{3\mu_0}{4\pi} \cdot (1 - 3\cos^2\beta) \cdot \frac{\vec{m}^2}{\vec{r}^2}. \tag{5}$$

For SPBs arranged in the experimentally observed cluster formations, two angles β were identified to be important for this theoretical investigation: β = 90° for SPBs placed within one vertical row and β ≈ 30° for SPBs that were ejected to the side of the original row formation (see Figure 4c). Assuming that SPBs are in physical contact with each other, i.e., the surface-to-surface distance $d$ is set to zero and, therefore, $|\vec{r}|$ = 2.8 µm, $|\vec{F}_{\text{dip}}|$ was calculated for β varying between 0° and 90°. The utilized magnetic moment $m_x$ of a single SPB was hereby obtained according to the calculations shown in the Supporting Information S2. The result, which is plotted in Figure 4d, suggests a critical angle of 54.7° where the sign of $F_{\text{dip}}$ is inverted, turning a repulsive interaction when coming from 90° towards an attractive interaction [40]. Consequently, SPBs arranged in the observed cluster formations experience, depending on their position, either repulsion or attraction, with the absolute magnitude of the dipolar force being higher for β = 30° as compared to β = 90°. This could hint at the physical reason for single SPBs being gradually ejected toward the side of an original row formation. It needs to be noted, however, that not all assumptions that were initially made hold true for ejected particles, e.g., their position is slightly shifted with respect to the DW center. Therefore, the presented calculations can only provide an estimate of the SPB dipolar interaction strengths.

Looking further at the remaining forces as a function of SPBs' surface-to-surface distance $d$, the results of the calculations are shown in Figure 4e. Hereby used parameters are listed in detail in the Supporting Information S2. The magnetic forces $F_{mag}(z)$ (red dashed line) and $F_{mag}(y)$ (blue solid line) were computed independently on $d$ according to equation (2) for a single SPB. The SPB was hereby assumed to reside at $z = 1600$ nm above the smallest DW at a position of maximum stray field gradient with respect to the $y$-direction. As already inferable from the simulated MFL (Figure 1c), the magnitude of the $z$-component along a hh/tt DW is significantly larger than the $y$-component at the corners of a magnetic stripe domain, consequently resulting in higher induced magnetic moments within the SPBs and a stronger magnetic force. Additionally, it can be deduced from the simulated MFL that $|H_z|$ and therefore also the magnetic force is larger for increasing stripe domain length, potentially explaining the SPB motion behavior observed for small external magnetic field pulse lengths, i.e., the particle valve functionality. For a critical pulse length, particles within the focal point, that are exposed to a reduced magnetic force, already transition to the non-linear transport regime while particles at larger stripe lengths with stronger magnetic forces are still linearly transportable. The clogging of SPBs can be physically interpreted as the occurrence of position-dependent critical transition time scales, which is an outstanding feature of the introduced domain pattern design. Returning to the magnetic forces connected to SPBs positioned within the magnetic pattern's focal point, $F_{mag}(z)$ is approximately 35 times larger than $F_{mag}(y)$, thus, promoting a preferred attraction of SPBs towards the positions of hh/tt DWs. Nonetheless, $F_{mag}(y)$ might still have a role in the stabilization of SPB cluster arrangements since particles sitting at the top and bottom of a formation are additionally attracted towards the positions of maximum $|H_y|$ at the corners of stripe domains. Both magnetic forces are, however, dominating the dipolar interaction between SPBs, signifying that both dipolar forces for $\beta = 30°$ (purple dashed line) and $\beta = 90°$ (green solid line) play a minor role in the physical formation of the observed particle clusters. Especially the repellent dipolar interaction for particles arranged within a vertical row, which usually suppresses the occurrence of SPB aggregates [10], seems to be overcome in this case by the stronger magnetic attraction towards a DW center. Additionally, the dipolar repulsion is weakened for the focal point of the domain pattern as compared to larger DW lengths, since stray field strengths are reduced in this case and a lower magnetic moment is consequently induced within the SPBs. A similar observation applies to the calculated electrostatic force $F_{el}$ (orange dash dotted line): Its magnitude only becomes larger than the magnetic force for very small surface-to-surface distances of the SPBs (d ⪅ 30 nm). Since $F_{el}$ is of repellent nature due to equal surface potentials of the SPBs, it is expected that particles cannot fully approach each other. This further hinders irreversible particle aggregation due to the attractive van der Waals force $F_{vdW}$ (pink dotted line), since for the considered range of $d$, the magnitude of $F_{el}$ is always larger than $F_{vdW}$.

*Conclusion*

In this work, a specifically designed magnetic parallel-stripe domain pattern was tailored as a functional LOC-element, combining the three functions particle "funneling", concentration, and "valve" for superparamagnetic beads (SPBs) in an aqueous medium. Three goals were set for the design of the element: 1.) Introducing a functionality to limit the maximum number of SPBs being transported above the pattern, 2.) bringing SPBs in close proximity to each other, thereby facilitating bead aggregation upon formation of molecular bridges, and 3.) creating an on/off switchable valve for directionally transported SPBs. The domain pattern was engineered to exhibit gradually increasing/decreasing stripe length in a periodic fashion while maintaining a constant stripe width of 5 μm. It was fabricated in an exchange biased thin film system with in-plane magnetization via ion bombardment induced magnetic patterning (IBMP). Magnetic force microscopy (MFM) imaging confirmed the occurrence of the stripe domains with opposite magnetization direction compared to the monodomain environment. Three types of domain walls (DWs) for the different magnetization configurations head-to-head (hh), tail-to-tail (tt), and side-by-side (ss) were identified, representing the sources for the magnetic stray field landscape (MFL) that is an essential component of the utilized particle transport concept. Spatial components of the MFL were simulated based on micromagnetic calculations for the domain pattern, revealing the field in $z$-direction above the center of a hh/tt DW to be the strongest, while additional field components in $y$-direction are expected as compared to previously investigated parallel-stripe domain patterns [10]. After adding an aqueous dispersion of Dynabeads M-270 (diameter of 2.8 μm) above the substrate, directed transport of the particles was initialized by applying external trapezoidal magnetic field pulses in $z$- and $x$-direction. As a striking feature of the observed motion behavior, formations of SPBs that started out as vertical rows were transformed into cluster formations with hexagonal-like arrangement of single particles upon reaching the sample area with smallest stripe domain length (focal point). This cluster formation has been reversibly turned back to the original row formation when leaving this focal point towards regions with increasing stripe domain length. In the present experiments, no more than 6 SPBs were found within a single cluster at the focal point of the domain pattern, demonstrating the possibility to produce particle aggregates with a defined number of particles. Past the focal point, these clusters turn back to particle rows each of which with this defined number of particles. This feature of the investigated functional element paves the way for a variety of applications, where a defined amount of magnetic particles is needed at a designated chip area. Studying the influence of the external magnetic field pulse length, a transition towards a blocked focal point was observed, i.e., passing of the SPBs through the focal point was inhibited by non-transportable particles stuck within the focal point. This showcases the particle "valve" and concentrator functionalities of the investigated transport system, with the external magnetic field pulse length being the lever to open or close the valve. Within a quantitative theoretical discussion, the main contributing forces on the SPBs were calculated to provide physical explanations for the observed particle clustering/de-clustering behavior. It was discovered that the magnetic force acting perpendicular to the substrate plane ($z$-direction) alongside occurring fringe fields at the upper/lower borders of magnetic stripe domains are most likely promoting the observed densely packed SPB cluster formations within the magnetic pattern's focal point. For larger stripe domain lengths, repulsive magnetostatic and electrostatic forces between single SPBs lead to the observed vertical row arrangements with approximately equal spacing between the particles. Combining the SPB motion concept presented within this work with adequately surface-functionalized particles, analyte detection based on induced irreversible particle aggregation is a promising application for LOC systems. The number of particles within one aggregate is hereby adjustable via the minimum stripe domain length as well as the particle size and concentration, making this "magnetophoretic funnel" especially potent for calibrating magnetoresistance-based sensing elements integrated into the chip substrate.

*Methods*

*Fabrication of magnetically patterned transport substrate*

The magnetic parallel-stripe domain pattern with gradually increasing/decreasing stripe length (stripe width of 5 μm) was obtained via IBMP [20, 23] of an EB thin film system. A $Cu^{5\ nm}/Ir_{17}Mn_{83}^{30\ nm}/Co_{70}Fe_{30}^{10\ nm}/Si^{20\ nm}$ layer stack was deposited onto a naturally oxidized Si (100) wafer piece (ca. 1 cm × 1 cm) by rf-sputtering at room temperature. Subsequently, the sample was subjected to a field cooling procedure to induce the in-plane direction of the EB. Therefore, the sample was annealed in a vacuum chamber (base pressure = 5 x $10^{-7}$ mbar) at 300 °C for 60 min in an in-plane magnetic field of 145 mT. For IBMP, a photoresist with a sufficient thickness to prevent 10 keV He ions from penetrating the magnetic layer system, was deposited on the sample surface via spin coating. The photoresist was structured to exhibit 5 μm wide stripe gaps with a 5 μm separation of adjacent stripes and a periodically repeating increase/decrease of the stripe length (see Figure S1 in the Supporting Information for image of the resulting resist structure). The stripe length was varied between 50 μm and 5 μm with an increment of 5 μm between adjacent stripes (except for the 50 μm and 5 μm long stripes which were repeated once before subsequent length modulation). The long axis of the stripes was positioned perpendicular to the initial EB direction (set by the field cooling procedure). Structuring of the resist was performed by utilizing direct laser writing lithography. Next, the sample was bombarded with a dose of 1 x $10^{15}$ $cm^{-2}$ He ions (kinetic energy of 10 keV) using a home-built Penning ion source.

For an antiparallel stripe magnetization with respect to the surrounding (protected by the resist layer), an in-plane homogenous magnetic field (100 mT) was applied antiparallel to the initial EB direction during ion bombardment. Afterward, the photoresist was removed by washing the sample several times with acetone. Then the surface was cleaned by rinsing the sample with acetone, isopropanol, and water. After drying, a 200 nm thick PMMA layer was deposited on top of the sample by spin-coating.

*Particle transport*

For inducing SPB motion, a home-built setup consisting of orthogonally placed Helmholtz coil pairs was used for the application of trapezoidal magnetic field pulses in *z*- and *x*-direction, i.e., perpendicular and parallel to the transport substrate plane. Each pulse consisted of a linear rising time for the magnetic field, a plateau time, and a linear drop time, with the pulse direction being periodically alternated between $H_{max}$ and $-H_{max}$. The duration of pulse rising and drop times was given by the pulse magnitude and the alteration rate of the external magnetic field ($3.2 \cdot 10^6$ Am$^{-1}$s$^{-1}$). The pulse magnitude was chosen to be $\mu_0 \cdot |H_{max,x}| = \mu_0 \cdot |H_{max,z}| = 1$ mT and a temporal phase shift of π/2 between pulse sequences in *z*- and *x*-direction was applied. Before initializing the transport experiment, a volume of 20 μL of a diluted dispersion of SPBs (Dynabeads M-270 Carboxylic Acid [41]) was pipetted into a microfluidic chamber adhered to the top of the magnetically patterned substrate. The chamber was fabricated by cutting a window of approximately 8 mm × 8 mm into a Parafilm sheet that is of the substrate's size. After sealing the chamber with a square-shaped glass coverslip, the transport substrate was placed in the middle of the Helmholtz coil arrangement with the substrate plane positioned perpendicular to the external *z*-field and in-plane domain magnetization direction positioned parallel to the external *x*-field. Approaching the sample with an optical bright field microscope (40× magnification objective,

N.A. = 0.6), SPB motion was recorded with an attached high speed camera (Optronis CR450x2) at a framerate of 1000 frames per second (fps) and an image resolution of 800 x 600 pixels.

*Micromagnetic simulations*

The simulation package MuMax3 [37] was utilized to compute the magnetization distribution $\vec{m}(x,y)$ within a region of interest (see black rectangle in Figure 1b) for the investigated stripe domain pattern with gradually modified stripe length. Two regions were defined for areas of the sample that were treated/untreated by He ion bombardment: The stripes themselves are ion bombarded and the surrounding environment is untreated. Depending on this categorization, differing magnetic properties were assigned: An exchange stiffness constant of $A_{\text{ex}} = 3 \cdot 10^{-11} \frac{\text{J}}{\text{m}}$ [42], a saturation magnetization of $M_S = 1.23 \cdot 10^6 \frac{\text{A}}{\text{m}}$ [43], a uniaxial anisotropy constant of $K = 4.5 \cdot 10^4 \frac{\text{J}}{\text{m}^3}$ [36] for the non-bombarded environment, and accordingly $A_{\text{ex}} = 3 \cdot 10^{-11} \frac{\text{J}}{\text{m}}$, $M_S = 1.18 \cdot 10^6 \frac{\text{A}}{\text{m}}$, $K = 3.375 \cdot 10^4 \frac{\text{J}}{\text{m}^3}$ for the bombarded stripes. Note that the values for saturation magnetization and anisotropy constant are slightly reduced for the bombarded region as uncovered by previous investigations [43, 44]. For implementing the exchange bias-related pinning of the respective domain magnetizations, additional biasing magnetic fields were defined for bombarded/non-bombarded areas with opposing directions. Here, the magnetic flux densities were chosen to be 13 mT for the non-bombarded regions and 6.7 mT for bombarded regions according to experimentally determined values from hysteresis loop measurements. The region of interest (20.48 μm × 40.96 μm) was discretized into cubic elements of 5 nm × 5 nm × 10 nm sizes and the simulation software computed the relaxed magnetization state of the described system, which was subsequently used for obtaining magnetic stray field components (see Figure 1c) via a dipole approximation.

AUTHOR INFORMATION

*Corresponding Author*

rico.huhnstock@physik.uni-kassel.de

*Author Contributions*

RH: Investigation, Formal analysis, Writing – original draft

LP: Investigation, Formal analysis, Writing – review & editing

MM: Formal analysis

PK: Investigation, Resources, Writing – review & editing

AE: Supervision, Conceptualization, Project administration, Funding acquisition, Writing – review & editing

*Funding Sources*

-

ACKNOWLEDGMENT

Fruitful discussions with Dr. Michael Vogel are gratefully acknowledged.

ABBREVIATIONS

SPB = superparamagnetic bead

LOC = Lab-on-a-chip

MFL = magnetic field landscape

EB = exchange bias

IBMP = ion bombardment induced magnetic patterning

hh = head-to-head

tt = tail-to-tail

DW = domain wall

MFM = magnetic force microscopy

PMMA = Poly(methyl methacrylate)

SUPPORTING INFORMATION

S1: Image of the utilized resist structure for IBMP

S2: Parameters for force calculations

REFERENCES

[1]  M. A. M. Gijs, "Magnetic bead handling on-chip: New opportunities for analytical applications," *Microfluidics and Nanofluidics,* vol. 1, no. 1, pp. 22-40, 2004.

[2]  N. Pamme, "Magnetism and microfluidics," *Lab on a Chip,* vol. 6, no. 1, pp. 24-38, 2006.

[3]  C. Ruffert, "Magnetic Bead-Magic Bullet," *Micromachines,* vol. 7, no. 2, p. 21, 2016.

[4]  L. J. Kricka, "Microchips, microarrays, biochips and nanochips: Personal laboratories for the 21st century," *Clinica Chimica Acta,* pp. 219-223, 2001.

[5]  A. Manz, N. Graber and H. Widmer, "Miniaturized total chemical analysis systems: A novel concept for chemical sensing," *Sensors and Actuators: B. Chemical,* vol. 1, pp. 244-248, 1990.

[6]  J. Knight, "Honey, I shrunk the lab," *Nature,* vol. 418, no. 6897, pp. 474-475, 2002.

[7]  A. Van Reenen, A. M. De Jong, J. M. Den Toonder and M. W. Prins, "Integrated lab-on-chip biosensing systems based on magnetic particle actuation-a comprehensive review," *Lab on a Chip,* vol. 14, no. 12, pp. 1966-1986, 2014.

[8]  C. Moerland, L. van IJzendoorn and M. Prins, "Rotating magnetic particles for lab-on-chip applications-a comprehensive review," *Lab on a Chip,* vol. 19, no. 6, pp. 919-933, 2019.

[9] S. Rampini, P. Li and G. Lee, "Micromagnet arrays enable precise manipulation of individual biological analyte–superparamagnetic bead complexes for separation and sensing," *Lab on a Chip,* vol. 16, no. 19, pp. 3645-3663, 2016.

[10] D. Holzinger, I. Koch, S. Burgard and A. Ehresmann, "Directed Magnetic Particle Transport above Artificial Magnetic Domains Due to Dynamic Magnetic Potential Energy Landscape Transformation," *ACS Nano,* vol. 9, no. 7, pp. 7323-7331, 2015.

[11] R. Abedini-Nassab, M. Pouryosef Miandoab and M. Sasmaz, "Microfluidic synthesis, control, and sensing of magnetic nanoparticles: A review," *Micromachines,* vol. 12, no. 7, p. 768, 2021.

[12] M. Reginka, H. Hoang, Ö. Efendi, M. Merkel, R. Huhnstock, D. Holzinger, K. Dingel, B. Sick, D. Bertinetti, F. W. Herberg and A. Ehresmann, "Transport Efficiency of Biofunctionalized Magnetic Particles Tailored by Surfactant Concentration," *Langmuir,* vol. 37, no. 28, pp. 8498-8507, 2021.

[13] R. Huhnstock, M. Reginka, A. Tomita, M. Merkel, K. Dingel, D. Holzinger, B. Sick, M. Vogel and A. Ehresmann, "Translatory and rotatory motion of exchange-bias capped Janus particles controlled by dynamic magnetic field landscapes," *Scientific Reports,* vol. 11, no. 1, p. 21794, 2021.

[14] A. Chen, T. Byvank, G. B. Vieira and R. Sooryakumar, "Magnetic microstructures for control of brownian motion and microparticle transport," *IEEE Transactions on Magnetics,* vol. 49, no. 1, pp. 300-308, 2013.

[15] A. Sarella, A. Torti, M. Donolato, M. Pancaldi and P. Vavassori, "Two-dimensional programmable manipulation of magnetic nanoparticles on-chip," *Advanced Materials,* vol. 26, no. 15, pp. 2384-2390, 2014.

[16] U. Sajjad, F. Klingbeil, F. Block, R. B. Holländer, S. Bhatti, E. Lage and J. McCord, "Efficient flowless separation of mixed microbead populations on periodic ferromagnetic surface structures," *Lab on a Chip,* vol. 21, no. 16, pp. 3174-3183, 2021.

[17] B. Lim, P. Vavassori, R. Sooryakumar and C. Kim, "Nano/micro-scale magnetophoretic devices for biomedical applications," *Journal of Physics D: Applied Physics,* vol. 50, no. 3, p. 033002, 2017.

[18] P. Tierno, F. Sagues, T. H. Johansen and T. M. Fischer, "Colloidal transport on magnetic garnet films," *Physical Chemistry Chemical Physics,* vol. 11, no. 42, pp. 9615-9625, 2009.

[19] A. Ehresmann, D. Lengemann, T. Weis, A. Albrecht, J. Langfahl-Klabes, F. Göllner and D. Engel, "Asymmetric magnetization reversal of stripe-patterned exchange bias layer systems for controlled magnetic particle transport," *Advanced Materials,* vol. 23, no. 46, pp. 5568-5573, 2011.

[20] A. Ehresmann, I. Koch and D. Holzinger, "Manipulation of superparamagnetic beads on patterned exchange-bias layer systems for biosensing applications," *Sensors (Switzerland),* vol. 15, no. 11, pp. 28854-28888, 2015.

[21] M. Urbaniak, M. Matczak, G. Chaves-O'Flynn, M. Reginka, A. Ehresmann and P. Kuświk, "Domain wall motion induced magnetophoresis in unpatterned perpendicular magnetic anisotropy Co layers with Dzyaloshinskii-Moriya interactions," *Journal of Magnetism and Magnetic Materials,* vol. 519, p. 167454, 2021.

[22] A. Mougin, S. Poppe, J. Fassbender, B. Hillebrands, G. Faini, U. Ebels, M. Jung, D. Engel, A. Ehresmann and H. Schmoranzer, "Magnetic micropatterning of FeNi/FeMn exchange bias bilayers by ion irradiation," *Journal of Applied Physics,* vol. 89, no. 11, pp. 6606-6608, 2001.

[23] A. Ehresmann, I. Krug, A. Kronenberger, A. Ehlers and D. Engel, "In-plane magnetic pattern separation in NiFe/NiO and Co/NiO exchange biased bilayers investigated by magnetic force microscopy," *Journal of Magnetism and Magnetic Materials,* vol. 280, no. 2-3, pp. 369-376, 2004.

[24] P. Kuświk, A. Ehresmann, M. Tekielak, B. Szymański, I. Sveklo, P. Mazalski, D. Engel, J. Kisielewski, D. Lengemann, M. Urbaniak, C. Schmidt, A. Maziewski and F. Stobiecki, "Colloidal domain lithography for regularly arranged artificial magnetic out-of-plane monodomains in Au/Co/Au layers," *Nanotechnology,* vol. 22, no. 9, p. 095302, 2011.

[25] E. Albisetti, D. Petti, M. Pancaldi, M. Madami, S. Tacchi, J. Curtis, W. P. King, A. Papp, G. Csaba, W. Porod, P. Vavassori, E. Riedo and R. Bertacco, "Nanopatterning reconfigurable magnetic landscapes via thermally assisted scanning probe lithography," *Nature Nanotechnology,* vol. 11, no. 6, pp. 545-551, 2016.

[26] I. Berthold, U. Löschner, J. Schille, R. Ebert and H. Exner, "Exchange bias realignment using a laser-based direct-write technique," *Physics Procedia,* vol. 56, no. C, pp. 1136-1142, 2014.

[27] T. Ueltzhöffer, R. Streubel, I. Koch, D. Holzinger, D. Makarov, O. G. Schmidt and A. Ehresmann, "Magnetically Patterned Rolled-Up Exchange Bias Tubes: A Paternoster for Superparamagnetic Beads," *ACS Nano,* vol. 10, no. 9, pp. 8491-8498, 2016.

[28] B. Lim, S. R. Torati, K. W. Kim, X. Hu, V. Reddy and C. G. Kim, "Concentric manipulation and monitoring of protein-loaded superparamagnetic cargo using magnetophoretic spider web," *NPG Asia Materials,* vol. 9, no. 3, pp. e369-e369, 2017.

[29] C. P. Gooneratne, I. Giouroudi, C. Liang and J. Kosel, "A giant magnetoresistance ring-sensor based microsystem for magnetic bead manipulation and detection," *Journal of Applied Physics,* vol. 109, no. 7, p. 07E517, 2011.

[30] C. Gooneratne, R. Kodzius, F. Li, I. Foulds and J. Kosel, "On-Chip Magnetic Bead Manipulation and Detection Using a Magnetoresistive Sensor-Based Micro-Chip: Design Considerations and Experimental Characterization," *Sensors,* vol. 16, no. 9, p. 1369, 2016.

[31] M. Urbaniak, D. Holzinger, A. Ehresmann and F. Stobiecki, "Magnetophoretic lensing by concentric topographic cylinders of perpendicular magnetic anisotropy multilayers," *Biomicrofluidics,* vol. 12, no. 4, p. 044117, 2018.

[32] P. Li, D. Gandhi, M. Mutas, Y. F. Ran, M. Carr, S. Rampini, W. Hall and G. U. Lee, "Direct identification of the herpes simplex virus: UL27 gene through single particle manipulation and optical detection using a micromagnetic array," *Nanoscale,* vol. 12, no. 5, pp. 3482-3490, 2020.

[33] R. Wirix-Speetjens, G. Reekmans, R. De Palma, C. Liu, W. Laureyn and G. Borghs, "Magnetoresistive biosensors based on active guiding of magnetic particles towards the sensing zone," *Sensors and Actuators, B: Chemical,* vol. 128, no. 1, pp. 1-4, 2007.

[34] S. Rampini, P. Li, D. Gandhi, M. Mutas, Y. F. Ran, M. Carr and G. U. Lee, "Design of micromagnetic arrays for on-chip separation of superparamagnetic bead aggregates and detection of a model protein and double-stranded DNA analytes," *Scientific Reports,* vol. 11, no. 1, p. 5302, 2021.

[35] Y. F. Ran, C. Fields, J. Muzard, V. Liauchuk, M. Carr, W. Hall and G. U. Lee, "Rapid, highly sensitive detection of herpes simplex virus-1 using multiple antigenic peptide-coated superparamagnetic beads," *Analyst,* vol. 139, no. 23, pp. 6126-6134, 2014.

[36] D. Holzinger, N. Zingsem, I. Koch, A. Gaul, M. Fohler, C. Schmidt and A. Ehresmann, "Tailored domain wall charges by individually set in-plane magnetic domains for magnetic field landscape design," *Journal of Applied Physics,* vol. 114, no. 1, p. 013908, 2013.

[37] A. Vansteenkiste, J. Leliaert, M. Dvornik, M. Helsen, F. Garcia-Sanchez and B. Van Waeyenberge, "The design and verification of MuMax3," *AIP Advances,* vol. 4, no. 10, p. 107133, 2014.

[38] W. Nolting, Grundkurs Theoretische Physik 3, Springer, 2013.

[39] J. N. Israelachvili, Intermolecular and surface forces, Elsevier, 2011.

[40] I. Koch, M. Langner, D. Holzinger, M. Merkel, M. Reginka, R. Huhnstock, A. Tomita, C. Jauregui Caballero, A. Greiner and A. Ehresmann, "3D Arrangement of Magnetic Particles in Thin Polymer Films Assisted by Magnetically Patterned Exchange Bias Layer Systems," *Particle & Particle Systems Characterization,* vol. 38, no. 12, p. 2100072, 2021.

[41] Thermofisher, "Dynabeads M-270 Carboxylic Acid," 13 March 2023. [Online]. Available: https://www.thermofisher.com.

[42] D. V. Berkov, C. T. Boone and I. N. Krivorotov, "Micromagnetic simulations of magnetization dynamics in a nanowire induced by a spin-polarized current injected via a point contact," *Physical Review B - Condensed Matter and Materials Physics,* vol. 83, no. 5, p. 54420, 2011.

[43] H. Huckfeldt, A. Gaul, N. D. Müglich, D. Holzinger, D. Nissen, M. Albrecht, D. Emmrich, A. Beyer, A. Gölzhäuser and A. Ehresmann, "Modification of the saturation magnetization of exchange bias thin film systems upon light-ion bombardment," *Journal of Physics: Condensed Matter,* vol. 29, no. 12, p. 125801, 2017.

[44] N. D. Müglich, M. Merkel, A. Gaul, M. Meyl, G. Götz, G. Reiss, T. Kuschel and A. Ehresmann, "Preferential weakening of rotational magnetic anisotropy by keV-He ion bombardment in polycrystalline exchange bias layer systems," *New Journal of Physics,* vol. 20, no. 5, p. 053018, 2018.

## S1 Image of utilized resist structure for IBMP

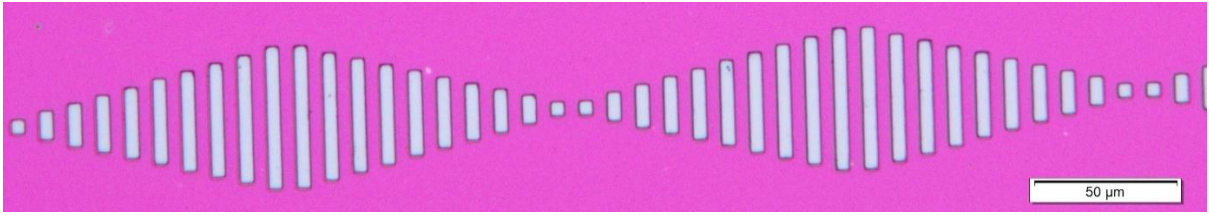

*Figure S1: Brightfield microscope image of the utilized resist structure for ion bombardment induced magnetic patterning, fabricated via direct laser writing lithography.*

## S2 Parameters for force calculations

For determining the magnetostatic forces, either generated by the magnetic field landscape (MFL) onto the superparamagnetic beads (SPBs) $F_{mag}$ or generated by dipole-dipole interaction between two SPBs $F_{dip}$, the absolute magnetic moment $m_{abs}$ of the particles, which correlates with the present magnetic field, was taken from literature[1]. Therefore, the absolute magnetic field was calculated according to

$$H_{abs} = \sqrt{H_{eff,x}^2 + H_{eff,y}^2 + H_{eff,z}^2},$$

with $H_{eff}$ resulting from the superposition $H_{eff} = H_{MFL} + H_{ext}$ of the MFL and the applied external magnetic field. $H_{MFL}$ was determined from the simulated results shown in Figure 1c of the manuscript for a SPB that is residing with its centre 1600 nm above the ferromagnetic layer of the thin film system and that is located centrally above a head-to-head (hh) domain wall for a magnetic stripe domain of 5 µm width and length. For $H_{ext}$, the experimentally applied magnitudes $H_{ext,x}\mu_0 = H_{ext,z}\mu_0 = 1\,\text{mT}$ were used, with $\mu_0$ being the vacuum magnetic permeability. Hence, absolute magnetic field values of 2591 A/m for the calculations of the magnetic force in z-direction $F_{mag}(z)$ and the dipolar force $F_{dip}(\beta)$ together with 2623 A/m for the calculation of the magnetic force in y-direction $F_{mag}(y)$ were retrieved. The latter is slightly higher than the former, as here the SPB was placed closer to the corner of a magnetic stripe domain, where the magnetic force $F_{mag}(y)$ was expected to be maximal. Consulting literature[1], $m_{abs}$ was determined to be 0.819 Am²/kg for the calculations of $F_{mag}(z)$ as well as $F_{dip}(\beta)$ and 0.858 Am²/kg for the calculation of $F_{mag}(y)$. For using the absolute magnetic moment of a single SPB, the respective mass was calculated from the density[1] $\rho = 1600\,\frac{\text{kg}}{\text{m}^3}$ and the volume $V = \frac{4}{3}\pi r^3$ with

r = 1.4 µm. Considering only the projection of the magnetic moment onto the *x*-direction for the calculation of $F_{dip}(\beta)$, $m_x$ was determined according to

$$m_x = m_{abs} \cdot \cos \alpha,$$

where $\alpha = \tan^{-1} \frac{H_{eff,z}}{H_{eff,x}}$. For computing the DLVO forces $F_{el}$ and $F_{vdW}$ according to equations (3) and (4) in the manuscript, the utilized literature values are summarized in the following table.

Table 1: Used parameters for the calculation of DLVO forces between two spherical particles of equal size

| | |
|---|---|
| Debye length $1/\kappa$ in water[2] | 100 nm |
| Zeta potential of Dynabead M-270[3] | -28.28 mV |
| Hamaker constant A for two polystyrene particles interacting with each other in water[4] | $1.4 \cdot 10^{-20}$ J |

It has to be noted that the surface potential $\psi_0$ required for the calculation of $F_{el}$ was approximated by the zeta potential of a single SPB found in literature.

(1) Grob, D. T.; Wise, N.; Oduwole, O.; Sheard, S. Magnetic Susceptibility Characterisation of Superparamagnetic Microspheres. *J. Magn. Magn. Mater.* **2018**, *452*, 134–140. https://doi.org/10.1016/j.jmmm.2017.12.007.

(2) Butt, H.; Graf, K.; Kappl, M. *Physics and Chemistry of Interfaces*; Wiley, 2003. https://doi.org/10.1002/3527602313.

(3) Wise, N.; Grob, D. T.; Morten, K.; Thompson, I.; Sheard, S. Magnetophoretic Velocities of Superparamagnetic Particles, Agglomerates and Complexes. *J. Magn. Magn. Mater.* **2015**, *384*, 328–334. https://doi.org/10.1016/j.jmmm.2015.02.031.

(4) Israelachvili, J. N. *Intermolecular and Surface Forces*, 3.; Academic Press: Burlington, 2011.